\def\be{\begin{equation}}
\def\te{\end{equation}}
\def\bea{\begin{eqnarray}}
\def\nn{\nonumber}
\def\tea{\end{eqnarray}}

\def\d{\delta}

\newskip\humongous \humongous=0pt plus 1000pt minus 1000pt

\newif\ifdtup

\documentclass[onecolumn,aps]{revtex4b5}
\usepackage{setspace}
\usepackage{graphicx}
\input{epsf}

\begin{document}
\title{Some asymptotic properties of duplication graphs}
\author{Alpan Raval} 
\email{araval@kgi.edu,alpan.raval@cgu.edu}
\affiliation{Keck Graduate Institute of Applied Life Sciences, \\535 Watson Drive, Claremont, CA 91711.\\
Department of Mathematics, Claremont Graduate University \\
710 N. College Avenue,
 Claremont, CA 91711.}

\begin{abstract}
{Duplication graphs are graphs that grow by duplication of existing vertices, and are important models of biological networks, including protein-protein interaction networks and gene regulatory networks. Three models of graph growth are studied: pure duplication growth, and two two-parameter models in which duplication forms one element of the growth dynamics. A power-law degree distribution is found to emerge in all three models. However, the parameter space of the latter two models is characterized by a range of parameter values for which duplication is the predominant mechanism of graph growth. For parameter values that lie in this ``duplication-dominated'' regime, it is shown that the degree distribution either approaches zero asymptotically, or approaches a non-zero power-law degree distribution very slowly. In either case, the approach to the true asymptotic degree distribution is characterized by a dependence of the scaling exponent on properties of the initial degree distribution. It is therefore conjectured that duplication-dominated, scale-free networks may contain identifiable remnants of their early structure. This feature is inherited from the idealized model of pure duplication growth, for which the exact finite-size degree distribution is found and its asymptotic properties studied.\\
PACS number(s): 87.23.kg, 05.40.-a, 02.50.Cw\\
Electronic address: {\tt araval@kgi.edu}, {\tt alpan.raval@cgu.edu}} 
\end{abstract}
\maketitle

\section{Introduction}
\label{sec:intro}

The study of evolving graphs as a means to describe the power-law degree distribution of large networks has become increasingly relevant in recent years, starting with the study of the preferential attachment model of graph growth \cite{bara} that models a diverse range of man-made and natural networks. Graphs that grow by duplication of existing vertices \cite{chung,kim,sole} are particularly relevant to the study of biological networks, including protein-protein interaction networks and genetic regulatory networks, because they mimic the process of gene duplication by duplication of vertices, i.e, by creation of new vertices that have exactly the same set of connections as pre-existing vertices in the graph. Various processes of graph growth in which duplication forms one element of the growth dynamics have been shown to exhibit scale-free behavior at late times, characterized by a power-law dependence of the degree distribution $p(k)$ of the graph, i.e., $p(k)\sim k^{\gamma}$, where $\gamma$ is the scaling exponent \cite{footnote1}. This has led to the notion that biological networks possess features in common with other well-studied, albeit disparate, networks, including the Internet and metabolic networks \cite{storgatz,albert}. A particularly attractive feature of such scale-free networks is their putative robustness and tolerance of error \cite{bara,banavar,albert2}.  

At the same time, it is not so well-known that graphs that grow predominantly by the duplication process have features that are distinct from other scale-free graphs. These features become particularly stark and revealing in the limit of pure duplication growth. One such feature is the lack of the ``self-averaging'' property \cite{kim}, i.e., the property that an individual realization of graph growth does not asymptotically reach the degree distribution of an \underline{ensemble} of such realizations. Specifically, it was shown \cite{chung} that the number of distinct ``orbits'' (the subsets of nodes that are connected to exactly the same sets of nodes) remains invariant under any one realization of pure duplication growth. Therefore the number of distinct degrees of the graph (where the degree of a node is defined as the number of its nearest neighbors) also remains invariant. This lack of ``self-averaging'' property may be formalised into an appropriate notion of lack of ergodicity in the graph dynamics.

Another distinct feature of duplication graphs is the lack of clear emergence of an asymptotic (long-time) solution for the degree distribution  of an ensemble of realizations \cite{footnote2}. While the dynamics of a single realization of the duplication process can be clarified \cite{chung} in terms of invariance in the number of orbits, the dynamics of an ensemble of such processes is quite non-trivial (because of lack of self-averaging) and is discussed below.  In a model proposed earlier \cite{sole} that includes duplication as well as mutation by edge removal and addition, a breakdown of the asymptotic stationary solution is found to occur in the analysis. For a range of parameters in which duplication is the dominant process of graph growth (the duplication-dominated regime), the analytically obtained stationary solution has negative average degree and the scaling exponent does not agree with that obtained from numerical simulations. Further analysis of the same model \cite{kim} reveals that the degree distribution at late times depends sensitively on initial conditions, although the dependence itself is not clarified. 

One of the common threads in the analysis of duplication graphs is the assumed existence of a non-trivial, asymptotic, stationary degree distribution. While the scale-free preferential attachment model \cite{bara} and other related models do have an asymptotic solution that is stationary, this is not generally true. For our purposes, we will define a stationary degree distribution to be a time-independent, \underline{non-zero} degree distribution \cite{footnote3}.   

A number of questions naturally emerge from the above observations. Some of them are:\\
(1) Do ensembles of duplication graphs (i.e., graphs in which the mechanism of growth is predominantly by means of duplication) have stationary asymptotic degree distributions?\\
(2) Do ensembles of duplication graphs exhibit asymptotically scale-free behavior?\\
(3) How does the asymptotic degree distribution depend on initial conditions?

In this work, these questions are first answered in the context of pure duplication growth, where, as is shown below, an exact solution for the degree distribution at all times can be obtained analytically. This is followed by a discussion of these issues in mixed models that contain duplication as a component of the dynamics. It is conjectured that duplication-dominated growth may serve to define a new class of models that are asymptotically non-stationary (or, at best, quasi-stationary) but nevertheless may exhibit scale-free behavior. In spite of their lack of asymptotic stationarity, these models could well describe realistic biological networks.

\section{Pure duplication growth}
Consider an undirected graph that grows by pure duplication. We will assume that the graph has $m_0$ vertices at time $t=0$, and that time progresses in units of $1$. At each time step, an existing vertex is picked at random and duplicated, i.e., a new vertex is added to the graph with the same set of edges as an existing vertex. The number of vertices therefore increases by one at each time step and the total number of vertices at time $t$ is $t+m_0$. Consequently, the maximum possible degree at time $t$ is $k_{\rm max}(t)=t+m_0-1$. As shown earlier \cite{chung}, any specific process of this type (i.e., a realization of this dynamics) leaves the number of orbits, and therefore, the number of distinct degrees in the graph, invariant. We will, however, consider the dynamics of an ensemble of such processes and denote the degree distribution of this ensemble by $p(k,t)\equiv$ the probability of finding a vertex of degree $k$ at time $t$.

Since every vertex has equal probability of being duplicated at a given time step, the probability $p_{\rm new}(k,t)$ that a new vertex has degree $k$ at time $t$ is given by
\begin{equation}
p_{\rm new}(k,t)=p(k,t-1).
\end{equation}
Furthermore, the probability $p_{\rm ndup}(k',t)$ that a vertex of degree $k'$ is a neighbor of a duplicating vertex is proportional to its degree. Demanding that a vertex of maximum degree is a neighbor of a duplicating vertex with probability $1$ then gives
\begin{equation}
p_{\rm ndup}(k,t)=k(m_0+t-1)^{-1}.
\end{equation}
From the above we can find the number of vertices of degree $k$ at time $t$ as
\begin{eqnarray}
n(k,t)&=&p_{\rm new}(k,t)(n(k,t-1)+1)+(1-p_{\rm new}(k,t))n(k,t-1) \nonumber \\
& &+p_{\rm ndup}(k-1,t)n(k-1,t-1)-p_{\rm ndup}(k,t)n(k,t-1),
\end{eqnarray}
where the first two terms on the right-hand-side (RHS) of the above equation describe the contribution of the duplicating vertex itself and distinguish between the two cases: (a) the duplicating vertex is of degree $k$, and (b) the duplicating vertex is not of degree $k$; the third term comes from vertices of degree $k-1$ increasing their degree because they are neighbors of a duplicating vertex, and the fourth term is a loss term for vertices that were of degree $k$ at the previous time and have since increased their degree due to neighbor duplication. 


Noting that $n(k,t)=p(k,t)(t+m_0)$, one can derive the following master equation for $p(k,t)$:
\begin{equation}\label{simpleeq}
p(k,t)-p(k,t-1)=\frac{k-1}{t+m_0}p(k-1,t-1)-\frac{k}{t+m_0}p(k,t-1).
\end{equation}
The above equation holds for all $k \geq 0$ (with $p(-1,t)=0$ for all $t$). However, the dynamics of isolated vertices (vertices of degree $0$) is decoupled from the dynamics of higher degree vertices. Indeed, one obtains $p(0,t)$ is constant for all time and $p(k,t)$ for $k \geq 1$ does not depend on $p(0,t)$. Because of this decoupling property, we will only consider solutions of Eq. (\ref{simpleeq}) for $k \geq 1$, supplemented by the equation $p(0,t)=p(0,0)$. Correspondingly, we will only consider graphs with a minimum degree of $1$, with the understanding that ensembles of graphs that contain isolated vertices can be subdivided into two ensembles, one ensemble of graphs whose minimum vertex degree is $1$, and another ensemble of graphs that only contain isolated vertices. The dynamics of these two ensembles is then decoupled and we may only consider the non-trivial dynamics of the ensemble with minimum vertex degree of $1$.


By inspection of Eq. (\ref{simpleeq}), a naive solution is obtained. This is a ``stationary'' solution with scaling exponent $\gamma=-1$ satisfying $kp(k)=(k-1)p(k-1)$, i.e., $p(k) \sim k^{-1}$. Note that this solution is not a global solution at any finite time because it is not correctly normalised: demanding $\sum_{k=1}^{k_{\rm max}(t)}p(k)=1$ causes the solution to be non-stationary, in which case it is not a solution at all. This solution can, at best, therefore be an asymptotic solution, and even so, hold only for finitely many values of $k$, because the sum of $k^{-1}$ over infinitely many values of $k$ is divergent, and the normalization condition would fail to hold. Indeed, from an analysis of the exact degree distribution below, we find that this stationary solution is not an asymptotic solution at all, although the pure duplication growth limit in earlier analyses \cite{kim,sole,chung} yields a scaling exponent of $-1$.



\subsection{Exact degree distribution for pure duplication growth}

It turns out that the master equation (\ref{simpleeq}) is simple enough to solve exactly in terms of the initial degree distribution $p(k,0)$. By writing out each term on the RHS of the master equation in terms of distributions at earlier times, one notices that $p(k,t)$ is a sum of terms of the general form
\begin{equation}
p(k-i,0)\frac{(k-1)(k-2)\cdots(k-i)(t+m_0-k)(t-1+m_0-k)\cdots(i+1+m_0-k)}{(t+m_0)(t+m_0-1)\cdots(m_0+1)},
\end{equation}
where $i$ runs from $0$ to $t$. Furthermore, there are $t!/(i!(t-i)!)$ terms of this type. Putting all this together, one obtains
\begin{eqnarray}
\lefteqn{p(k,t)=\sum_{i=0}^{t}p(k-i,0)\frac{t!}{i!(t-i)!}\times}\nonumber \\
& &\frac{(k-1)\cdots(k-i)(t+m_0-k)\cdots(i+1+m_0-k)}{(t+m_0)(t+m_0-1)\cdots(m_0+1)}.
\end{eqnarray}
Changing the dummy variable $i$ to $j=k-i$ and noting that, in the initial distribution, the minimum degree is $1$ while the maximum degree is $m_0-1$, one finally obtains, after some simplification
\begin{equation}
\label{exact}
p(k,t) = \left( \begin{array}{c}
t+m_0 \\
m_0
\end{array} \right)^{-1}\sum_{j={\rm max}(k-t,1)}^{{\rm min}(k,m_0-1)}\,
\left( \begin{array}{c}
t+m_0-k \\
m_0-j
\end{array} \right)
\left( \begin{array}{c}
k-1  \\
j-1
\end{array} \right)\, p(j,0)
\end{equation}
In the sum above, it is understood that for values of $k$ and $t$ such that the lower limit of the sum is larger than the upper limit, $p(k,t)=0$. The above solution corresponds to a mixture, via the initial distribution, of a hypergeometric distribution \cite{angus} and may be readily verified by direct substitution into Eq. (\ref{simpleeq}).

\subsection{Asymptotic analysis}

The exact degree distribution, Eq. (\ref{exact}) above, shows that, for $t \gg m_0$, there are three regimes of $k$ values for which the degree distribution has potentially qualitatively different behavior. The first regime is $1\leq k < m_0-1$, for which only terms from $j=1$ up to $j=k$ contribute in the sum. The second regime is $m_0-1 \leq k \leq t+1$, for which the entire support of the initial degree distribution contributes to the sum (i.e., all terms from $j=1$ to $j=m_0-1$). The third regime is $t+1 < k < k_{\rm max}(t) \equiv t+m_0-1$, for which only terms from $j=k-t$ up to $j=m_0-1$ contribute to the sum. At late times, the number of distinct $k$ values in the second regime ($t+m_0-1$ values) is much larger than the number of distinct $k$ values in the first and third regimes. We will therefore restrict our analysis to values of $k$ that correspond to the second regime. For the asymptotic analysis below, we will further assume that $m_0 \ll k \ll t$. 
 
In order to study the late-time behavior of the degree distribution, the asymptotic expansion of the Gamma function \cite{gradshteyn} is used to obtain the following asymptotic results, valid for $m_0 \ll k \ll t$ and $1\leq j \leq m_0-1$:
\be
\left(\begin{array}{c}
t+m_0 \\
m_0
\end{array} \right) \sim  \frac{t^{m_0}}{m_0!}\left(1+O(t^{-1})\right),
\te
\be
\left(\begin{array}{c}
t+m_0-k  \\
m_0-j
 \end{array} \right) \sim \frac{t^{m_0-j}}{(m_0-j)!}\left(1+O(t^{-1})\right),
\te
\be
\left(\begin{array}{c}
k-1  \\
j-1
\end{array} \right) \sim  \frac{k^{j-1}}{(j-1)!}\left(1+O(k^{-1})\right). 
\te
Substituting these into Eq. (\ref{exact}), one obtains, for $m_0 \ll k \ll t$,
\be
\label{asymp1}
p(k,t) \sim \frac{m_0}{t} \sum_{j=1}^{m_0-1}\left(\begin{array}{c}
m_0-1 \\
j-1
\end{array}\right)\,\left(\frac{k}{t}\right)^{j-1}\,p(j,0)\,\,\left(1+O(k^{-1})\right).
\te
Since every successive term in the sum above is multiplied by an additional factor of $k/t \ll 1$, the dominant contribution to $p(k,t)$ comes from the lowest non-zero value of $j$ such that $p(j,0) \neq 0$. This value of $j$ is the lowest non-zero degree in the initial ensemble of graphs. Defining $k_{\rm min}$ as the lowest non-zero degree in the initial distribution, we obtain the approximate asymptotic result
\be
\label{relntomin}    
p(k,t) \sim \frac{m}{t} \left(\begin{array}{c}
m_0-1 \\
k_{\rm min}-1
\end{array}\right)\,\left(\frac{k}{t}\right)^{k_{\rm min}-1}\,p(k_{\rm min},0)\,\,\left(1+O(k^{-1})\right).
\te
It follows that the asymptotic degree distribution approaches zero as $t^{-k_{\rm min}}$ for large $t$ and is therefore non-stationary. However, for large, finite $t$, the following result is obtained.\\

\noindent \underline{ The asymptotic degree distribution for pure duplication graphs, although non-stationary,}\\
\underline{has a scaling exponent of $\gamma = k_{\rm min} -1$, where $k_{\rm min}$ is the smallest non-zero} \\
\underline{degree in the initial graph.}\\

In particular, the scaling exponent is \underline{positive} when $k_{\rm min}>1$. [This behavior does not cause any normalization problems as $t \rightarrow \infty$ because $p(k,t) \rightarrow 0$ in this limit]. Figure 1 gives plots of the asymptotic degree distribution generated from numerical simulations of the master equation. For the case $k_{\rm min}=1$, it is found that the degree distribution is uniform, while for $k_{\rm min}=2$, $p(k)$ has a linear dependence on $k$, consistent with the above result.

For realistic graphs, such as most biological networks of interest, it is usually the case that $k_{\rm min} =1$. If these graphs evolved by means of a pure duplication process, the late-time degree distribution of an ensemble of such graphs would be dominated by a uniform distribution. 
We now examine the features of the asymptotic degree distribution that are amenable to a direct analysis of the master equation.

\subsection{Direct asymptotic analysis}

\noindent The asymptotic behavior of the degree distribution obtained so far relies on knowledge of the exact solution (\ref{exact}). It is of interest to know what features of the asymptotic degree distribution can be obtained directly from the master equation, without recourse to the exact solution. This is especially important in the analysis of more complex models, where the exact degree distribution for all time and for all values of $k$ may be analytically intractable.

We first note that the lack of existence of a stationary asymptotic degree distribution may be deduced immediately from a generating function \cite{gardiner} approach to the problem. Assuming that a stationary asymptotic degree distribution exists, with $p(k,t)=p(k,t-1)\equiv p(k)$ for all $k \geq 1$, and defining the generating function $\phi(x)$,
\be
\label{gener}
\phi(x)=\sum_{k=1}^{\infty} x^k\,p(k),
\te
one obtains from the master equation (\ref{simpleeq}) the following equation for $\phi(x)$:
\be
x(x-1)\frac{d\phi}{dx}=0,
\te
which gives $\phi(x)=$ constant for $0<x<1$. [The normalization condition $\phi(1)=1$ then implies that the constant equals $1$]. This is inconsistent with the fact that $p(k)\neq 0$ for some $k\geq 1$. Hence the assumption of a stationary distribution leads to a contradiction and therefore a stationary distribution cannot exist.\\

In order to analyse the non-stationary asymptotic distribution, one may assume that the asymptotic degree distribution is of the separable form:
\be
\label{separ}
p(k,t) \sim \sum_c g_c(t)\,f_c(k),
\te
where the possible values of $c$ are to be determined. Since the master equation is a linear homogenous equation, one may further demand that every term in the above sum satisfies the master equation. [Note that the true asymptotic solution (\ref{asymp1}) is indeed of the form (\ref{separ}).] A typical term in the sum above can then be substituted into the master equation (\ref{simpleeq}). After some rearrangement of terms, one obtains
\be
(t+m_0)\left(1-\frac{g_c(t)}{g_c(t-1)}\right)=k-(k-1)\frac{f_c(k-1)}{f_c(k)}.
\te
Since the LHS of the above equation is a function of $t$ alone and the RHS a function of $k$ alone, each side must be separately constant, leading to the pair of equations
\begin{eqnarray}
\label{eq1}
(t+m_0)\left(1-\frac{g_c(t)}{g_c(t-1)}\right)&=& c,\\
\label{eq2}
k-(k-1)\frac{f_c(k-1)}{f_c(k)} &=& c.
\end{eqnarray}
Equation (\ref{eq1}) above gives rise to divergent growth in $g_c(t)$ if $c<0$. A requirement is therefore $c>0$ (the $c=0$ case corresponds to a stationary solution, which has already been eliminated), a condition on the allowed values of $c$. With this condition, Eqs. (\ref{eq1}) and (\ref{eq2}) are readily solved to yield 
\begin{eqnarray}
\label{eq1sol}
g_c(t)&=&g_c(0)\,\frac{\Gamma(m_0+1)}{\Gamma(m_0-c+1)}\,\frac{\Gamma(t+m_0-c+1)}{\Gamma(t+m_0+1)} \sim t^{-c} \\
\label{eq2sol}
f_c(k)&=& f_c(1)\Gamma(2-c)\,\frac{\Gamma(k)}{\Gamma(k-c+1)} \sim k^{c-1}
\end{eqnarray}
Therefore, a power-law degree distribution with exponent $\gamma = c-1$ is obtained. From Eq. (\ref{eq1sol}), the lowest possible value of $c$ will dominate the late-time behavior. Note that, although $c$ is as yet undetermined, the analysis establishes the correct relationship between the exponent characterizing the rate at which the degree distribution falls to zero ($-c$) and the scaling exponent ($c-1$). This is evident by comparison of Eqs. (\ref{eq1sol}) and (\ref{eq2sol}) to Eq. (\ref{relntomin}).  This relationship is a testable one. As shown in the next section, a similar relationship can be derived from the asymptotic analysis of a more complex model.

In order to obtain the allowed values of the constant $c$ by a direct asymptotic analysis, we resort to an eigenvalue method that is described in the Appendix. The method shows that the allowed values of $c$ are the positive integers, $c=n$, $n=1,2,\ldots$, consistent with the exact solution (\ref{exact}). The lowest possible value of $c$ is then $c=1$, giving rise to a uniform degree distribution at late times. We are therefore able to capture most features of the exact solution by a direct asymptotic analysis, the missing feature being the relationship between the initial degree distribution and the lowest value of $c$.

\section{A duplication-mutation model}

We now consider a more general, two-parameter growth model suggested earlier \cite{sole} as a model for the evolutionary growth of the proteome that involves both duplication and mutation events. The model includes pure duplication growth as a special case. Assuming that the initial graph has $m_0$ nodes, the graph evolves according to the following rules: (i) a vertex is selected at random and duplicated, (ii) the links emanating from the newly generated vertex are removed with probability $\delta$, and (iii) new links are created between the new vertex and all other vertices with probability $\beta/(t+m_0-1)$ (where $t+m_0$ are the total number of vertices in the graph at time $t$). The processes of link addition and removal are necessarily correlated. However, for $\delta \ll 1$, it is reasonable to approximate the evolution by uncorrelated addition and removal \cite{sole}. With this assumption, the master equation for $p(k,t)$ is 
\begin{eqnarray}
p(k,t)-p(k,t-1)&=&\frac{(k+1)\d}{t+m_0}\,p(k+1,t-1)-\frac{k+2\beta}{t+m_0}\,p(k,t-1) \nn \\
\label{solemaster}
& &+\,\frac{(1-\d)(k-1)+2\beta}{t+m_0}\,p(k-1,t-1).
\end{eqnarray} 
Although the above equation describes the duplication-mutation process only for $\d \ll 1$, the equation is still a valid master equation for all values of $\d$ and will be studied for all values of $\d$ first before focussing on the duplication-dominated regime $\d \ll 1$. The equation that describes the duplication-mutation process for all values of $\d$, a further generalization of Eq. (\ref{solemaster}) above, has also been derived \cite{sole} and its asymptotic behavior for $\d > 1/2$ has been studied in detail \cite{kim}. The eventual case of interest here is $\d <1/2$. Thus, Eq. (\ref{solemaster}) will be sufficient for our purposes. Note that the limiting case $\d= 0$, $\beta=0$ corresponds to pure duplication growth. 

\subsection{Condition for an asymptotically stationary degree distribution}

To obtain the condition for the existence of an asymptotic stationary distribution, we assume a stationary normalizable distribution to begin with, proceed with the analysis, and search for a contradiction for some range of parameters. Indeed, setting $p(k,t)=p(k,t-1)=p(k)$ in Eq. (\ref{solemaster}), one obtains,
\be
(k+1)\d \,p(k+1) - (k+2\beta)\,p(k)+\left((1-\d)(k-1)+2\beta\right)\,p(k-1)=0.
\te
The corresponding generating function $\phi(x)$ is given by
\be
\label{genser}
\phi(x)=\sum_k\,x^k\,p(k).
\te
Before analysing the equation satisfied by the generating function, it is important to note that, for $p(k)$ to be a normalizable probability distribution, the series (\ref{genser}) must converge in the limit $x \rightarrow 1$. Furthermore, $\phi(x)$ must be analytic at $x=0$. With this in mind, we turn to the equation satisfied by $\phi(x)$:
\be
\left((1-\d)x-\d\right)\frac{d\phi}{dx}+2\beta \phi=0.
\te
The solution to the above equation is 
\be
\phi(x)=a \d^{-2\beta/(1-\d)} \mid 1 - x(\d^{-1}-1) \mid^{-2\beta/(1-\d)},
\te
where $a$ is an integration constant.
The above expression is analytic at $x=0$ and can be expanded in a Taylor series in powers of $x$ to obtain the probabilities $p(k)$. However, the Taylor series converges only if
\be
\mid x \mid (\d^{-1}-1)<1.
\te
Demanding that the series converge as $x \rightarrow 1$ then yields the condition
$\d > 1/2$. For $\d \leq 1/2$, the series is divergent \cite{footnote4}, which contradicts the assumption of a stationary normalizable probablility distribution $p(k)$. Therefore, we find that, for $\d \leq 1/2$, the asymptotic distribution is not stationary \cite{footnote5}.

\subsection{Asymptotic degree distribution for $\d \leq 1/2$}

\noindent Since the asymptotic distribution is not stationary for $\d \leq 1/2$, we may now consider solutions of (\ref{solemaster}) of the separable form
\be
p(k,t)\sim f_c(k)\,g_c(t).
\te
As in the previous section, these solutions are labeled by the separation constant $c$. One then obtains the pair of equations
\begin{eqnarray}
\label{mutt}
(t+m_0)\left(1-\frac{g_c(t)}{g_c(t-1)}\right) &=& c\\
\label{mutk}
(k+1)\d \,f_c(k+1) - (k+2\beta -c)\,f_c(k)+\left((1-\d)(k-1)+2\beta\right)f_c(k-1)&=&0.
\end{eqnarray}
It is clear from Eq. (\ref{mutt}) that one must have $c \geq 0$ for $g_c(t)$ to remain bounded as $t \rightarrow \infty$. Since $c=0$ corresponds to the stationary case, we will restrict attention to $c>0$. First, one finds from Eq. (\ref{mutt}),
\be
g_c(t)=g_c(0)\,\frac{\Gamma(t+m-c+1)\Gamma(m+1)}{\Gamma(t+m+1)\Gamma(m-c+1)} \sim t^{-c},
\te
as $t \rightarrow \infty$.

While the full asymptotic solution for $f_c(k)$ is difficult to obtain from Eq. (\ref{mutk}), we may carry out a Taylor expansion of $f_c(k+1)$ and $f_c(k)$ for large values of $k$, i.e.,
\begin{eqnarray}
f_c(k+1)&\simeq& f_c(k)+ \frac{df_c}{dk} \\
f_c(k-1)&\simeq& f_c(k)-\frac{df_c}{dk}
\end{eqnarray}
After substituting the above in Eq. (\ref{mutk}) and solving the resulting first order differential equation, one finds
\be
\gamma = \frac{c}{1-2\d}-1,
\te
resulting in the asymptotic ($t \rightarrow \infty$) solution
\be
p(k,t) \sim t^{-c}\,k^{c/(1-2\d)-1},~~~k \gg 1.
\te  
Again, the separation of variables analysis of the asymptotic degree distribution does not fix the allowed values of $c$. The eigenvalue method outlined in the Appendix, gives, to first order in $\delta$,
\be
c = 2\beta(1-\delta)+n(1-2\delta) + O(\delta^2),~~~n=0,1,2,\ldots
\te
Therefore, the late time solution will be dominated by the lowest value of $c$ that is consistent with initial conditions. The lowest possible such value is $c=-2\beta(1-\delta)$. It should be emphasized, however, that the above range of values of $c$ is only valid for $\delta \ll 1$. 

Figure 2 displays a plot of the degree distribution when $\delta=\beta=0.1$. In this case, the lowest possible value of $c$ is $c=0.18$, giving rise to an analytically predicted scaling exponent $\gamma \simeq -0.775$. Direct simulation of the master equation, shown in Fig. 2, gives approximate power law behavior with a scaling exponent of about $-0.73$, in reasonable agreement with the analytical result.

It may be argued that duplication-dominated growth ($\d < 1/2$) in this model is unrealistic because the mean degree $\langle k \rangle_t$ grows without bound \cite{sole,kim}, whereas realistic, large, biological networks have small mean degree. This argument is, however, unfounded. For the duplication-mutation model, it has been shown \cite{sole,kim} that $\langle k \rangle_t \sim t^{1-2\d}$ for large $t$ and for $\d < 1/2$. Therefore, if $\d$ is less than but sufficiently close to $1/2$ the mean degree will grow very slowly and remain small even when the size of the graph is large. Thus, $\d < 1/2$ could well be a viable region of parameter space, although, as shown here, the analysis of graph growth would require that the assumption of asymptotic stationary behavior be discarded. If the lowest allowed value of $c$ does depend on initial conditions (as in the pure duplication case), large biological networks may contain important clues about the structure of such networks very early in evolution. 

\section{A model with duplication and preferential attachment}

We now consider another two-parameter model of graph growth that also contains pure duplication growth as a special case but for which an asymptotic stationary distribution always exists everywhere in parameter space \underline{except} at the point corresponding to pure duplication growth. It will be seen that, although an asymptotic stationary distribution exists, the actual degree distribution approaches its stationary value very slowly in the duplication-dominated regime. Therefore, even at late times (corresponding to large graphs), the degree distribution is more accurately described by a \underline{quasi-stationary} distribution (in a manner clarified below) rather than by the true asymptotic stationary distribution. This model therefore serves to identify another possible feature of duplication-dominated growth, namely, quasi-stationary behavior, that may well hold in other, more realistic descriptions.

The growth model is a combination of pure duplication growth, and growth by simple scale-free, preferential attachment \cite{bara}. We start with an initial graph at time $t=0$ with $m_0$ vertices. At each time step \underline{one} of the following two processes can occur:\\
(a) An arbitrary vertex in the graph is duplicated (all vertices have equal probability of duplication), as in the pure duplication growth model, or\\
(b) A new vertex with $m$ edges is added to the graph. These edges are preferentially attached to the high-degree vertices, i.e., the probability that an old vertex will be linked to the new one is proportional to its degree.\\
We assume that process (a) occurs with probability $p_d$ and process (b) occurs with probability $1-p_d$. The model therefore has two parameters, $m$ and $p_d$. The case $p_d=1$ corresponds to pure duplication growth, while the case $p_d=0$ corresponds to growth by preferential attachment alone.

The master equation for such a growth model is a simple combination of the pure duplication and the scale-free preferential attachment master equations,
\begin{eqnarray}
\lefteqn{p(k,t)-p(k,t-1)=\frac{p_d}{t+m_0}\left\{(k-1)p(k-1,t-1)-kp(k,t-1)\right\}} \nonumber \\
& &+\frac{1-p_d}{t+m_0}\left\{\vphantom{\frac{m}{k}}\delta_{k,m}-p(k,t-1)\right.\nonumber \\\label{basic}
& &+\left.\frac{m}{\langle k \rangle_{t-1}}\left((k-1)p(k-1,t-1)-kp(k,t-1)\right)\right\},
\end{eqnarray}
where 
\begin{equation}
\langle k \rangle_{t-1}=\sum_k kp(k,t-1)
\end{equation}
is the mean degree at time $t-1$, and $\d _{k,m}$ is the Kronecker delta function.

\subsection{Existence of an asymptotic stationary distribution}

\noindent As before, we assume the existence of a stationary solution of Eq. (\ref{basic}) and check whether the generating function $\phi(x)$ is analytic at $x=0$ and whether the series converges as $x \rightarrow 1$. Assuming $p(k,t-1)=p(k,t)=p(k)$ in the limit $t \rightarrow \infty$, one obtains for $p(k)$
\be
\label{stateq}
\left(p_d + \frac{m(1-p_d)}{\langle k \rangle_{\infty}}\right)\left((k-1)p(k-1)-kp(k)\right)=(1-p_d)(p(k)-\d_{k,m}),
\te
where
\be
\langle k \rangle_{\infty}={\rm lim}_{t \rightarrow \infty}\langle k \rangle_t.
\te
The corresponding equation for the generating function $\phi(x)$, for $p_d \neq 1$, is
\be
\label{mygen}
\frac{x(1-x)}{\mu}\frac{d\phi}{dx}-\phi+x^m=0,
\te
where 
\be
\mu^{-1}=\frac{p_d}{1-p_d}+\frac{m}{\langle k \rangle_{\infty}} > 0.
\te
Equation (\ref{mygen}) can be solved to yield, after some simplification and a variable change,
\be
\phi(x)=\mu (1-x)^\mu\,x^m\,\int_0^1 \, ds\, s^{\mu +m-1}(1-xs)^{-\mu -1}+a\left(\frac{1-x}{x}\right)^\mu,
\te
where $a$ is an integration constant. Note that the radius of convergence of the Taylor expansion of $(1-x)^\mu$ is $1$ and that the radius of convergence of the Taylor expansion of $(1-xs)^{-\mu-1}$ is $1/s>1$. Furthermore, every term in the Taylor expansion of $(1-xs)^{-\mu-1}$ can be integrated to give a finite result, provided $m \neq 0$. Thus $\phi(x)$ is analytic at $x=0$, provided $a=0$ and $m \neq 0$. We therefore set the integration constant $a=0$. To show that the Taylor expansion converges at $x=1$, it is not enough to know that the Taylor expansion about $x=0$ has a radius of convergence of $1$. We further need to show that the integral over $s$ gives a finite result at $x=1$. 

In fact, the integral is divergent at $x=1$ for any $\mu \geq 0$. However, the factor of $(1-x)^\mu$ outside the integrand tends to $0$ as $x \rightarrow 1$. A more careful analysis is therefore required. To do this, we change variables from $s$ to $s'=(1-xs)/(1-x)$ and rewrite $\phi(x)$ in the form
\be
\phi(x)=\mu x^{-\mu}\int_1^{(1-x)^{-1}} ds' \,s'^{-\mu-1}\,\left(1-s'(1-x)\right)^{\mu+m-1}.
\te
Setting $x=1$ in the above yields $\phi(1)=1$, as required by normalization.  

We therefore find, for all $0\leq p_d <1$ and $m>0$, that the asymptotic distribution is stationary for this type of growth. However, to find the stationary distribution and the corresponding scaling exponent, we need to find $\langle k \rangle_{\infty}$, the asymptotic mean degree.

\subsection{Asymptotic mean degree}

\noindent The recursion equation for the evolution of the mean degree can be obtained by multiplying both sides of Eq. (\ref{basic}) by $k$ and summing over $k$. One obtains
\be
\label{meanrecur}
\langle k \rangle_t = \langle k \rangle_{t-1}\left(1+\frac{2p_d-1}{t+m_0}\right)+\frac{2m(1-p_d)}{t+m_0}.
\end{equation}

For $p_d < 1/2$, the above recusion gives rise to a finite asymptotic mean degree:
\be
\label{meanless}
\langle k \rangle_{\infty}=\frac{2m(1-p_d)}{1-2p_d},~~~p_d<1/2.
\te
For $p_d \geq 1/2$, the mean degree grows without bound as $t \rightarrow \infty$. To see this, we propagate Eq. (\ref{meanrecur}) back to $t=0$, giving
\begin{equation}
\label{kav}
\langle k \rangle_t = \frac{\Gamma (t+m_0 +2p_d)}{\Gamma(t+m_0+1)}\left\{\langle k \rangle_0 \frac{\Gamma(m_0+1)}{\Gamma(m_0+2p_d)} + 2m(1-p_d)\sum_{i=1}^t \frac{\Gamma(i+m_0)}{\Gamma(i+m_0+2p_d)}\right\}.
\end{equation}
The cases $p_d=1/2$ and $p_d > 1/2$ are considered separately. For $p_d=1/2$, the above equation simplifies to give
\be
\langle k \rangle_t = \langle k \rangle_0 + m \sum_{i=1}^t (m_0+i)^{-1}, ~~~p_d=1/2.
\te
For large $t$, one obtains the asymptotic behavior \cite{gradshteyn}
\be
\label{half}
\langle k \rangle_t = m\,\ln t + \langle k \rangle_0 -m\sum_{j=1}^{m_0}j^{-1}+m{\mathbf C}+ O(t^{-1}),~~~p_d=1/2,
\te
where ${\mathbf C}$ is Euler's constant. Thus the mean degree for $p_d=1/2$ grows logarithmically to infinity as $t \rightarrow \infty$.

For $p_d > 1/2$ (the duplication-dominated regime in this model), the sum over $i$ in Eq. (\ref{kav}) can be explicitly performed by expressing the ratio of Gamma functions in the sum in terms of the Beta function. Using an integral representation of the Beta function \cite{gradshteyn}, and interchanging the sum and the integral, one finds,
\begin{eqnarray}
\langle k \rangle_t &=& \frac{\Gamma (t+m_0 +2p_d)}{\Gamma(t+m_0+1)}\,\frac{\Gamma(m_0+1)}{\Gamma(m_0+2p_d)}\,\left\{\langle k \rangle_0 + \frac{2m(1-p_d)}{2p_d-1} \right\} - \frac{2m(1-p_d)}{2p_d-1} \\
\label{morethanhalf}
&\sim & t^{2p_d-1}\,\frac{\Gamma(m_0+1)}{\Gamma(m_0+2p_d)}\,\left\{\langle k \rangle_0 + \frac{2m(1-p_d)}{2p_d-1} \right\} - \frac{2m(1-p_d)}{2p_d-1},
\end{eqnarray}
where the second equation above holds for large $t$. Again, one finds for $p_d > 1/2$, that the mean degree grows without bound as a positive power of $t$ for large $t$.

We thus find that $\langle k \rangle_{\infty}=\infty$ for $p_d \geq 1/2$. Combining this result with the result (\ref{meanless}) for $p_d < 1/2$, we obtain
\begin{eqnarray}
\label{mu1}
\mu &=& 2(1-p_d), ~~~p_d<1/2,\\
\label{mu2}
&=& p_d^{-1}-1, ~~~p_d \geq 1/2.
\end{eqnarray}

\subsection{Asymptotic stationary distribution and quasi-stationary correction}

\noindent In order to obtain the asymptotic stationary distribution, we may directly solve the recursion of Eq. (\ref{stateq}) for $k > m$. One finds
\be
p(k)=p(m)\,\frac{\Gamma(m+\mu+1)}{\Gamma(m)}\,\frac{\Gamma(k)}{\Gamma(k+\mu+1)} \sim p(m)\,\frac{\Gamma(m+\mu+1)}{\Gamma(m)}\,k^{-\mu-1},
\te
where the last expression holds for $k \gg m$. A scale-free, stationary distribution therefore emerges, with scaling exponent $\gamma = -\mu-1$, and $\mu$ given by Eqs. (\ref{mu1}) and (\ref{mu2}) above. Note that this result breaks down in the pure duplication limit $p_d=1$, because in this limit the asymptotic distribution is \underline{not} stationary, as discussed earlier.

Although the above result for the scaling exponent is correct for infinitely large graphs, the scaling exponent for large but finite graphs may not even agree approximately with the asymptotic scaling exponent. To see this, note that the asymptotic scaling exponent $\gamma = -\mu-1$ was obtained by substituting the value of the mean degree at $t=\infty$ into the definition of $\mu$. For $p_d<1/2$ this mean degree is finite and it is expected that, as the graph grows, the mean degree will quickly approach its asymptotic value. However, in the duplication-dominated regime, $p_d \geq 1/2$, the asymptotic mean degree is infinite and therefore never approached, even if the graph is large. A simple example is the case $p_d=1/2$, for which the mean degree grows logarithmically with the size of the graph and may therefore be small even for large, finite graphs. Therefore, for values of $p_d$ greater than or equal to but close to $1/2$, it may be a better approximation to replace $\mu$ (and therefore $\gamma$) by its time-dependent value (obtained from the time-dependence of $\langle k \rangle_t$). This corresponds to a \underline{quasi-stationary} correction to the asymptotic stationary distribution, applied for large but finite graphs. 

Specifically, in the quasi-stationary regime, we have $p(k,t) \sim k^{\gamma(t)}$ with $\gamma(t)=-\mu(t)-1$ and
\be
\label{timemu}
\mu(t)^{-1} = \frac{p_d}{1-p_d} +\frac{m}{\langle k \rangle_t},
\te   
where, for large but finite graphs, $\langle k \rangle_t$ is given  by Eq. (\ref{half}) for $p_d=1/2$ and by Eq. (\ref{morethanhalf}) for $p_d > 1/2$. The scaling exponent therefore slowly drifts towards its true asymptotic value as the graph grows larger.

The effect of the quasi-stationary correction is studied in Figure 3 for the case $p_d=1/2$ and $m=6$. The graph is grown to approximately $1000$ vertices. In this case, the scaling exponent at $t=\infty$ is $\gamma=-2$, while the quasi-stationary correction gives $\langle k \rangle_{1000}\simeq 31.07$, $\mu^{-1} \simeq 1.19$, and a scaling exponent $\gamma \simeq -1.84$. This is in better agreement with the actual scaling exponent of about $-1.8$ obtained from the plot than the value $-2$.

\section{Discussion}

The asymptotic degree distributions in three models for graph growth have been analysed in this article: growth by pure duplication, and two two-parameter models in which duplication forms one element of growth. While pure duplication growth may be an unrealistic mechanism for a number of reasons (including lack of ergodicity, linear growth of the mean degree with the size of the graph, etc.), it serves as a useful idealized test case for the study of qualitative features, such as asymptotic non-stationarity and sensitivity to initial conditions, that may be present in more complex, more realistic models. By analysis of the exact degree distribution in the pure duplication model, we find that the asymptotic degree distribution of an ensemble of graphs subject to pure duplication growth is indeed non-stationary but nevertheless exhibits power-law behavior with a non-negative exponent that depends on initial conditions in a simple way -- the power law exponent is related to the lowest non-zero degree in the initial graph. The nature of the asymptotic degree distribution is also found from a direct asymptotic analysis of the master equation characterizing pure duplication growth, although such an analysis, being valid only in the asymptotic regime, does not relate the scaling exponent to initial conditions.  

The lack of existence of a stationary degree distribution is also found to occur in the duplication-dominated regime ($\delta \leq 1/2$) of the duplication-mutation model. For this model, $\delta = 1/2$ defines a critical boundary in parameter space that separates non-stationary and stationary asymptotic behavior. This also happens to be the critical boundary separating finite asymptotic mean degree and infinite asymptotic mean degree \cite{sole}. It is argued that, if $\delta$ is less than but sufficiently close to $1/2$, such a model could still describe realistic graphs, because the mean degree would increase very slowly with the size of the graph. The non-stationary asymptotic behavior of such duplication-dominated graphs could well depend on initial conditions in a manner similar to the pure duplication case, via the lowest allowed value of the constant $c$ that is consistent with initial conditions. 

For the model containing duplication growth combined with preferential attachment, an asymptotic stationary distribution is found to exist for all $p_d <1$. However, for the duplication-dominated regime, $p_d \geq 1/2$ (the critical boundary separating finite asymptotic mean degree and infinite asymptotic mean degree), the asymptotic degree distribution is more realistically described by a quasi-stationary distribution that takes into account the fact that the mean degree is always finite for large but finite graphs. $p_d=1/2$ can then be interpreted as a critical boundary separating stationary and quasi-stationary asymptotic degree distributions. On both sides of the critical boundary, the degree distribution has power-law behavior.

These results suggest that duplication-dominated graph growth may serve to model a new class of large networks whose degree distributions, although displaying power-law behavior, are not well-approximated by stationary distributions, even when these networks have large size. Based on the models studied, we have found that at least two kinds of non-stationary asymptotic behavior can occur in such networks: (a) one in which the probabilities drift to zero while the scaling exponent remains invariant as long as the network is large enough (non-stationary behavior), and (b) one in which the probabilities eventually converge to a non-zero, power-law distribution but the scaling exponent drifts slowly to its asymptotic value (quasi-stationary behavior). We also find that the scaling exponent will depend on initial conditions in both cases: in the non-stationary case, this dependence occurs via the allowed lowest value of the separation constant $c$, while in the quasi-stationary case, the scaling exponent depends on the mean degree in the initial graph, via Eq. (\ref{timemu}). Thus, duplication-dominated, scale-free networks may well contain early, and possibly identifiable, evolutionary remnants. 

We leave open to future work the question of the relationship, if any, between asymptotic stationarity of the degree distribution and ergodicity in the graph dynamics. \\

{\bf \center Acknowledgements}\\

The author thanks John Angus for pointing out an error in the original version of the manuscript and Greg Dewey, David Galas and Ashish Bhan for the benefit of numerous discussions on duplication graphs.


\newpage
\noindent {\bf Appendix: An eigenvalue method for analysing the time dependence of the degree distribution}\\

\noindent Consider the duplication-mutation model of Section 3. At late times ($t \gg m_0$), Eq. (\ref{solemaster}) can be expressed approximately as a differential equation in the time variable:
\be
\label{contin}
\frac{d{\mathbf p}(t)}{d(\ln t)}={\mathbf A}{\mathbf p}(t),
\te
where ${\mathbf p}(t)$ is a $t$-dimensional vector representation of the degree distribution, \\
${\mathbf p}(t)=[p(0,t)~p(1,t)~p(2,t)~\ldots ~p(t-1,t)]$, and the $t \times t$ matrix ${\bf A}$ is given by 
\be
{\mathbf A}=\left[ \begin{array}{rrrrrrr}
-2\beta & \delta & 0 & 0 & 0 & 0 & \cdots \\
2\beta & -(1+2\beta) & 2\delta & 0 & 0 & 0 & \cdots \\
0 & 1-\delta + 2\beta & -(2+2\beta) & 3\delta & 0 & 0 & \cdots \\
0 & 0 & 2(1-\delta)+2\beta & -(3+2\beta) & 4\delta & 0 & \cdots \\
\cdot & \cdot & \cdot & \cdot & \cdot & \cdot & \cdots \\
\cdot & \cdot & \cdot & \cdot & \cdot & \cdot & \cdots \\
\cdot & \cdot & \cdot & \cdot & \cdot & \cdot & \cdots 
\end{array} \right]
\te
The general solution to Eq. (\ref{contin}) is 
\be
{\mathbf p}(t)=\sum_{n=0}^{t-1} {\mathbf q}^{(n)}\,t^{\lambda_n},
\te
where $\{\lambda_n \}$ are the eigenvalues of the matrix ${\mathbf A}$ and the time-independent vectors ${\mathbf q}^{(n)}$ depend on the eigenvectors of ${\mathbf A}$ and on the  initial degree distribution. Note that the above solution justifies the separation-of-variables assumption made in Sections 2 and 3.

 In order to obtain the time-dependence of the degree distribution, we are interested in the eigenvalue spectrum of ${\mathbf A}$. While it is difficult to obtain the eigenvalues of ${\mathbf A}$ in general, it is quite straightforward to obtain them to leading order in $\delta$. Indeed, when $\delta =0$, the eigenvalue equation $\det ({\mathbf A}-\lambda {\mathbf I})=0$ immediately yields the eigenvalues (denoted by $\lambda_n^{(0)}$)
\be
\lambda_n^{(0)}=-n-2\beta, ~~~ n=0,1,2,\ldots
\te
For $\d \neq 0$, one finds, to first order in $\delta$,
\begin{eqnarray}
\det ({\mathbf A}-\lambda {\mathbf I})&=&(-1)^{t}\left\{\prod_{n=0}^{t-1}\,(n+2\beta + \lambda)\right.\nonumber\\
& & - \left.\sum_{n=1}^{t-1}\,n\delta\,\left(\prod_{l=0}^{n-2}(l+2\beta+\lambda)\right)(n-1+2\beta)\left(\prod_{l=n+1}^{t-1}(l+2\beta + \lambda)\right)\right\}\nonumber \\
& &+ O(\delta^2)
\end{eqnarray}
By solving for the eigenvalues to first order in $\delta$, one obtains
\be
\lambda_n=-n(1-2\delta)-2\beta(1-\delta)+O(\delta^2).
\te
The above gives the allowed values of the constant $c$ in Section 3, $c=-\lambda_n$. At late times, the degree distribution is dominated by the largest $\lambda_n$ (lowest $c$), obtained by setting $n=0$, as $\lambda_0=-c=-2\beta(1-\delta)$.

The results of the pure duplication growth described in Section 2 may be obtained by setting $\d =0, \beta=0$ in the above and removing the first row and first column of the matrix ${\mathbf A}$ (corresponding to decoupling the dynamics of isolated vertices from non-isolated ones). Removal of the first row and column is equivalent to discarding the eigenvalue $\lambda=0$. Denoting the remaining eigenvalues by $\lambda_n^{(0,0)}$, we then have
\be
\lambda_n^{(0,0)}=-n, n=1,2,\ldots
\te
The largest possible eigenvalue is then $-1$, resulting in $c=1$ and a uniform degree distribution as argued in Section 2.
\newpage


\newpage

\begin{figure}[hbt]
\centering
\leavevmode
\epsfysize=5.0in\epsffile{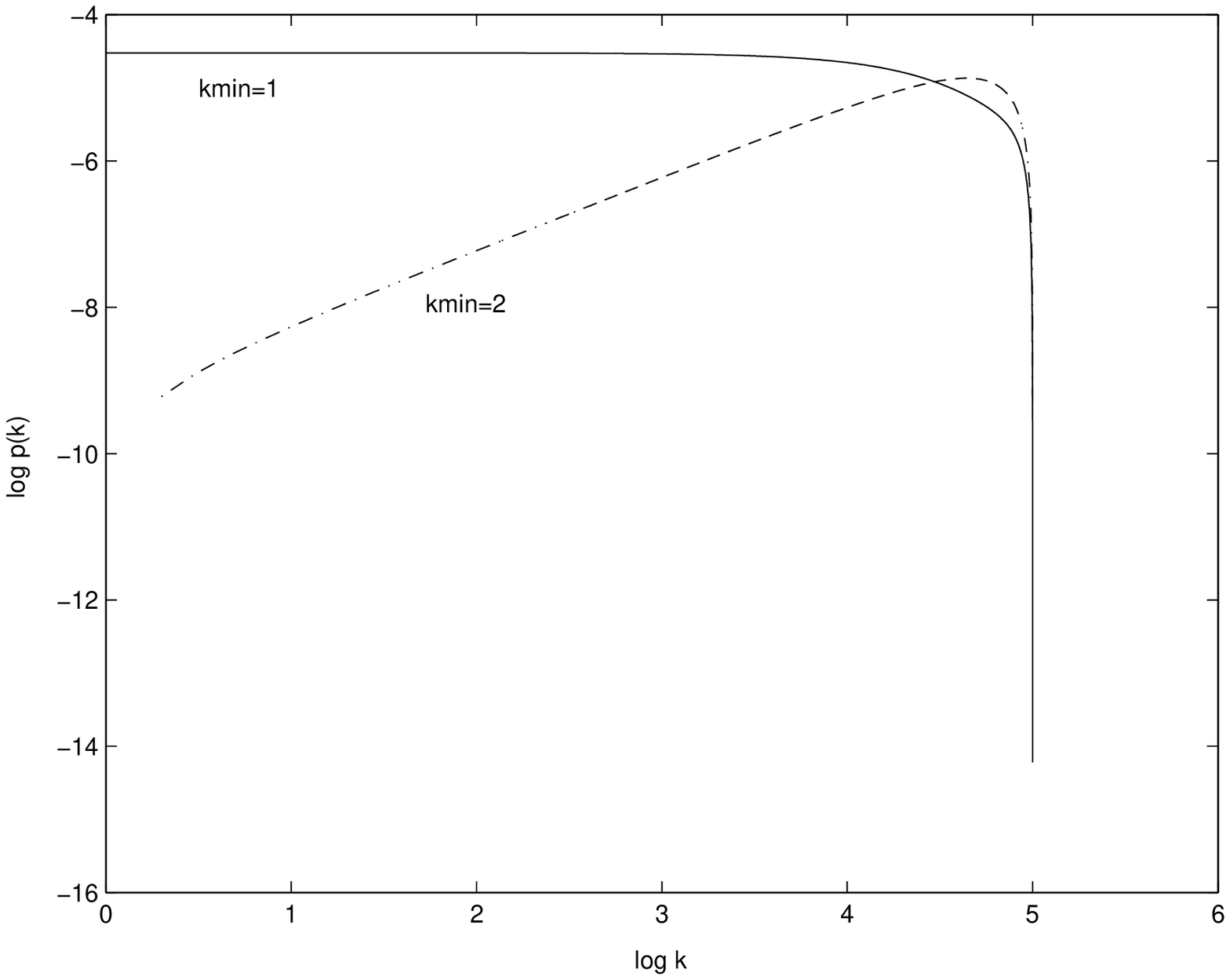}
\caption{\underline{Pure duplication growth}. The asymptotic degree distribution for an ensemble of graphs subject to pure duplication growth, obtained by simulation of the master equation. The solid line is generated by an initial ensemble in which the lowest non-zero degree is 1. The dashed line is generated by an initial ensemble in which the lowest non-zero degree is 2. The simulation was carried out to $10^5$ time steps.}
\end{figure}
\pagebreak

\begin{figure}[hbt]
\centering
\leavevmode
\epsfysize=5.0in\epsffile{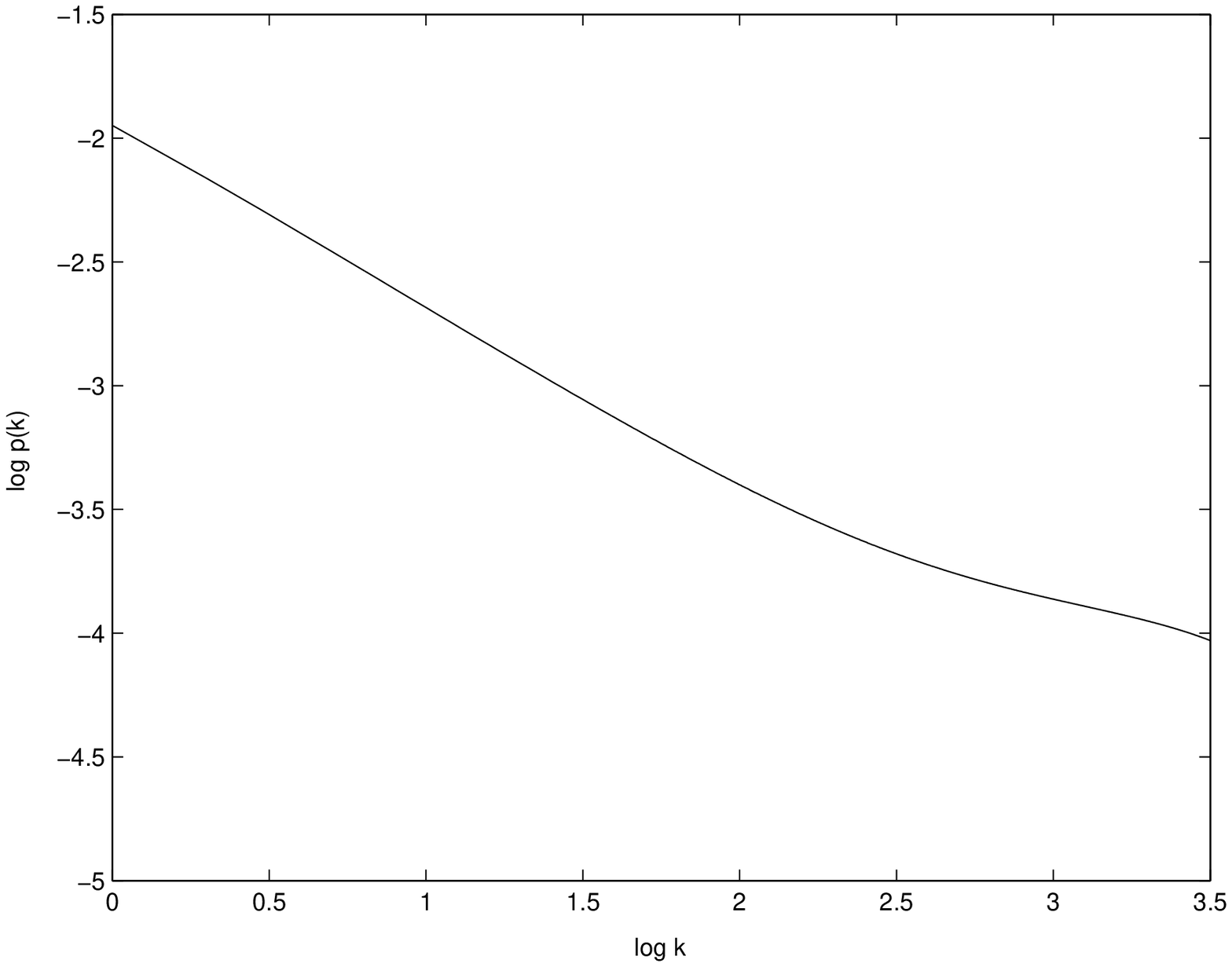}
\caption{\underline{Duplication-mutation growth}. The asymptotic degree distribution for an ensemble of graphs subject to duplication-mutation growth with parameters $\delta=\beta=0.1$. The scaling exponent is about $-0.72$, in good agreement with analytical predictions. The simulation was carried out to $10^6$ time steps.}
\end{figure}
\pagebreak

\begin{figure}[hbt]
\centering
\leavevmode
\epsfysize=5.0in\epsffile{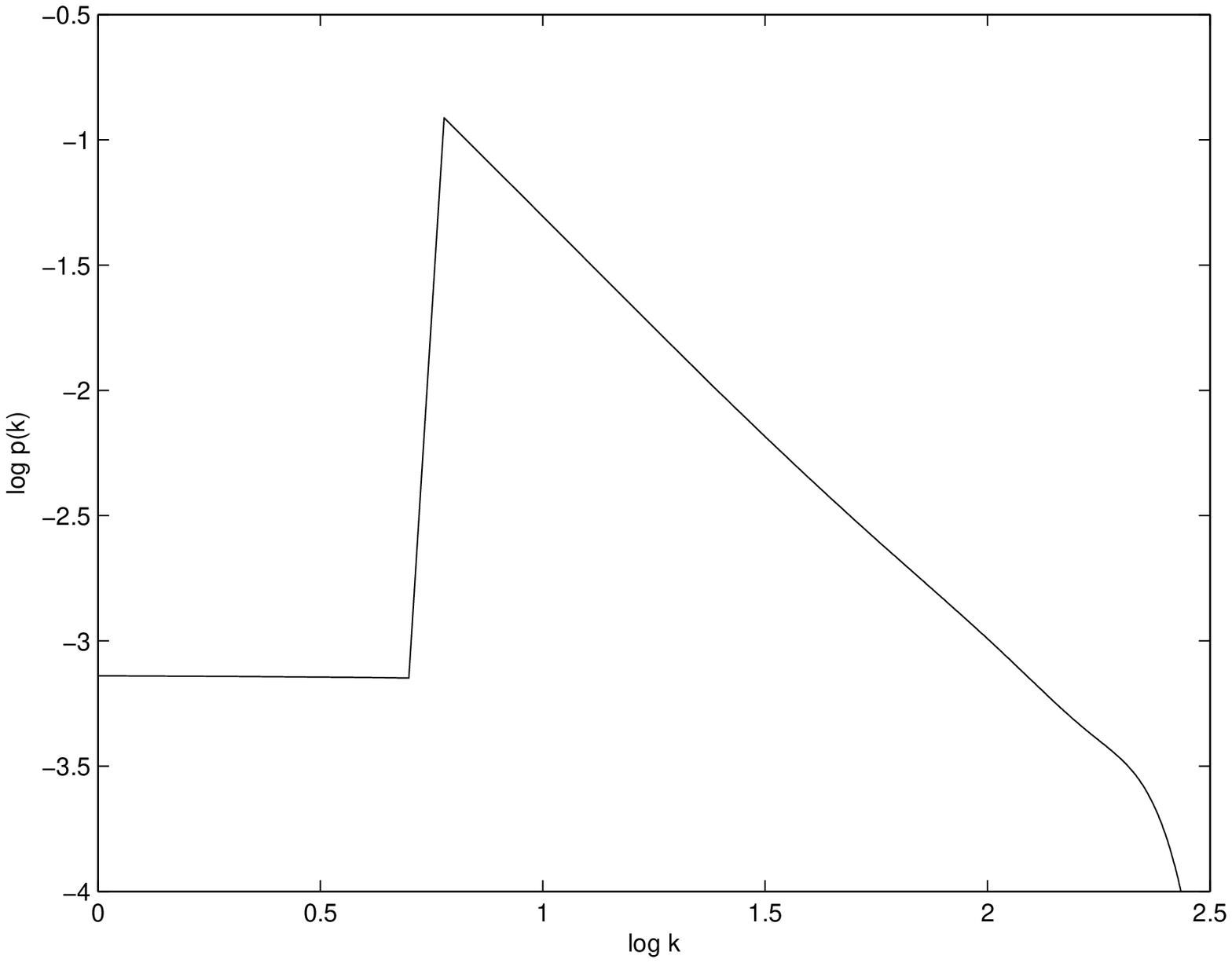}
\caption{\underline{Growth by duplication plus preferential attachment}. The asymptotic degree distribution for an ensemble of graphs subject to duplication plus preferential attachment with $\delta=1/2$ and $m=6$. The simulation was carried out only to $1000$ time steps in order to display the effect of the quasi-stationary correction. The true stationary asymptotic solution for this system has a scaling exponent of $-2$, while the quasi-stationary correction predicts a scaling exponent of about $-1.84$, in closer agreement with the actual value of about $-1.8$ obtained from the above simulation. The jump discontinuity at $k=6$ appears because the master equation is discontinuous at that value.}
\end{figure}


\begin{thebibliography}{999}
\bibitem{bara} A.-L. Barab\'{a}si and R. Albert, Science {\bf 286}, 509 (1999). 

\bibitem{chung} F. Chung, L. Lu, T. G. Dewey, and D. J. Galas, Journal of Computational Biology,  in press (2003).

\bibitem{kim} J. Kim, P. L. Krapivsky, B. Kahng, and S. Redner, Phys. Rev. E {\bf 66}, 055101(R) (2002).

\bibitem{sole} R. V. Sol\'{e}, R. Pastor-Satorras, E. D. Smith, and T. Kepler, Adv. Complex Syst. {\bf 5}, 43 (2002).

\bibitem{footnote1} Note that the scaling exponent as defined here differs from the usual definition by a minus sign.

\bibitem{storgatz} S. H. Storgatz, Nature (London) 410, {\bf 268} (2001).

\bibitem{albert} R. Albert and A.-L. Barab\'{a}si, Rev. Mod. Phys. {\bf 74}, 47 (2002).

\bibitem{banavar} J. R. Banavar, A. Maritan, and A. Rinaldo, Nature (London) {\bf 399}, 130 (1999).

\bibitem{albert2} R. Albert, H. Jeong, and A.-L. Barab\'{a}si,  Nature (London) {\bf 406}, 378 (2000).

\bibitem{footnote2} Note that the distinction between the dynamics of a single realization and that of an ensemble is necessary only for models that exhibit a lack of ``self-averaging''. For models such as the scale-free preferential attachment model \cite{bara}, this distinction is not an issue.

\bibitem{footnote3} An asymptotic probability distribution that is zero everywhere is possible if the random variable in question (here, the degree) has infinite range.

\bibitem{angus} J. Angus, private communication. If $Y$ is a hypergeometric random variable, i.e., $Y \sim {\rm Hypergeometric}(m_0-1,k-1,t+m_0-1)$, then the distribution in Eq. (\ref{exact}) can be written as $p(k,t)=\left(m_0/(t+m_0)\right)\,\langle p(Y+1,0) \,\theta({\rm min}(k-1,m_0-1)-Y) \rangle$, where $\theta(x)$ is the usual Heaviside function, and $\langle \,\rangle$ denotes expected value with respect to the hypergeometric distribution.  

\bibitem{gradshteyn} I. S. Gradshteyn and I. M. Rhyzhik, Table of Integrals, Series and Products. Academic Press, Inc. (1980).

\bibitem{gardiner} C. W. Gardiner, Handbook of stochastic methods, second ed. Springer (1985).

\bibitem{footnote4} Although the series expansion of $\mid 1 - x(\d^{-1}-1) \mid^{-2\beta/(1-\d)}$ is divergent at $x=1$ for $\d \leq 1/2$, $\phi(1)$ is still finite. However, this is not sufficient to guarantee normalizability of the probability distribution. The meaning of this inconsistency is that the expansion (\ref{genser}) breaks down at $x=1$ for $\d \leq 1/2$.

\bibitem{footnote5} It has been shown \cite{sole} that, for $\d \leq 1/2$, the mean degree grows without bound as $t \rightarrow \infty$ instead of approaching a finite value. While for this model, the unbounded growth of the mean degree coincides with the breakdown of stationarity, this is not generally true. The next section discusses a model where the mean degree grows without bound, yet the asymptotic degree distribution is stationary.
\end{thebibliography}
\end{document}